\documentclass[acmlarge]{acmart}

\usepackage{amssymb}

\AtBeginDocument{%
  }

\usepackage{algorithm}
\usepackage{algpseudocode}

\usepackage{tikz}
\usetikzlibrary{arrows.meta, positioning, shapes.geometric, fit}
\usepackage[most]{tcolorbox}

\usepackage{graphicx}
\usepackage{subcaption}
\usepackage{tabularx}
 \usepackage{amsmath,amssymb,mathtools}
 \usepackage[most]{tcolorbox}
\usepackage{array}

\setcitestyle{nosort}

\begin{document}
\title{Architectural Classification of XR Workloads: Cross-Layer Archetypes and Implications}

\author{Xinyu Shi}

\author{Simei Yang}

\author{Francky Catthoor}

\renewcommand{\shortauthors}{Shi et al.}

\begin{abstract}
Edge and mobile platforms for augmented and virtual reality—collectively referred to as extended reality (XR) must deliver deterministic ultra-low-latency performance under stringent power and area constraints. However, the diversity of XR workloads is rapidly increasing, characterized by heterogeneous operator types and complex dataflow structures. This trend poses significant challenges to conventional accelerator architectures centered around convolutional neural networks (CNNs), resulting in diminishing returns for traditional compute-centric optimization strategies. Despite the importance of this problem, a systematic architectural understanding of the full XR pipeline remains lacking.
In this paper, we present an architectural classification of XR workloads using a cross-layer methodology that integrates model-based high-level design space exploration (DSE) with empirical profiling on commercial GPU and CPU hardware. By analyzing a representative set of workloads spanning 12 distinct XR kernels, we distill their complex architectural characteristics into a small set of cross-layer workload archetypes (e.g., capacity-limited and overhead-sensitive). Building on these archetypes, we further extract key architectural insights and provide actionable design guidelines for next-generation XR SoCs. Our study highlights that XR architecture design must shift from generic resource scaling toward phase-aware scheduling and elastic resource allocation in order to achieve greater energy efficiency and high performance in future XR systems.
\end{abstract}



\maketitle

\section{Introduction}

In recent years, the increasing diversity and complexity of augmented and virtual reality (AR/VR) applications have driven increased demand for device platforms that can efficiently support these workloads~\cite{ILLIXR, Vasarainen2021}. Edge and mobile XR (extended reality) platforms, in particular, are required to maintain deterministic, ultra-low end-to-end latency and high frame rates while operating within stringent constraints on power, area, and thermal budgets~\cite{VR-Energy}. Unlike cloud-based AI services, edge and mobile XR systems must continuously process high-resolution sensor data and generate immersive graphics on-device in real time. As XR software stacks evolve from isolated vision kernels to increasingly heterogeneous on-device workloads that combine multiple perception and rendering stages, XR computation graphs become substantially larger and more diverse. This evolution exposes a fundamental challenge: the efficiency of fixed-function, CNN-centric accelerators diminishes when confronted with diverse operator types and dataflows.

Existing accelerators are largely optimized for dense convolutions and GEMM with regular tensor shapes, and CNN-centric designs explicitly tailor dataflow and buffering to exploit structured reuse (e.g., Eyeriss)~\cite{eyeriss}. Prior work has shown that rigid, highly specialized mappings (e.g., systolic-style execution) can suffer substantial efficiency loss under irregular or sparse computation patterns due to unfavorable data mappings and reduced utilization~\cite{sigma,MAERI}. In contrast, XR workloads frequently incorporate a broader mix of primitives beyond convolution/GEMM, including transforms, resampling, and feature matching, which commonly introduce data-dependent indexing, irregular dependencies, and intermediate tensors that must be explicitly stored and reused across stages~\cite{ILLIXR}. Collectively, these properties increase non-sequential memory traffic and inflate the working set, so the dominant bottleneck can shift across stages—e.g., from compute underutilization to bandwidth pressure, and then to on-chip capacity limits or synchronization/scheduling overheads. As a result, scaling peak compute throughput or bandwidth alone is often insufficient to consistently improve performance and energy efficiency for XR workloads.

Despite the acuteness of this hardware–software mismatch, a systematic architectural understanding of XR workloads remains limited. Prior studies often focus on isolated kernels (e.g., cost-volume, pose estimation), task-specific accelerators, or high-level algorithmic abstractions that obscure microarchitectural behavior and mapping-induced effects. What is still missing is a comprehensive, evidence-driven characterization that links the diversity of XR operators to concrete architectural sensitivities—such as on-chip storage residency, memory bandwidth demand, and control or synchronization overheads—so that architects can reason about XR systems beyond CNN-centric assumptions.

In this work, we address this gap by presenting an architectural classification of XR workloads. We adopt a cross-layer methodology that bridges empirical profiling on commercial hardware and model-based analytical design space exploration (DSE). On the profiling side, we combine GPU kernel-level profiling with CPU ISA-level instrumentation to capture both operator-level utilization and end-to-end application behaviors that cannot be inferred from GPU kernels alone, such as control-flow intensity and memory-access regularity. On the modeling side, we develop a lightweight analytical DSE framework to quantify energy–latency trade-offs as a function of on-chip memory capacity and a set of representative mapping choices, including tiling, buffering, and kernel fusion. To support consistent analysis across heterogeneous operators, we adopt a reconfigurable near-memory SIMD-style compute template as an analytical abstraction, rather than as a concrete microarchitectural proposal. By triangulating profiling metrics with analytical sweeps, we identify recurring bottleneck transitions—from compute-bound to bandwidth-bound, and from bandwidth-bound to capacity- or overhead-bound—under systematic changes in on-chip capacity and mapping. These transitions define a small set of cross-layer workload archetypes that capture common performance-limiting mechanisms across XR workloads.

The primary contributions of this work are as follows.
\vspace{-0.2em}
\begin{itemize}
    \item \textbf{Comprehensive architectural characterization.} We provide an in-depth analysis of 12 representative XR workloads, quantifying their sensitivity to compute throughput, memory hierarchy behavior, and execution overhead using both empirical profiling and analytical DSE.
    \item \textbf{Identification of cross-layer workload archetypes.} We formalize recurring behavioral patterns shared across XR workloads, enabling a compact abstraction for reasoning about architectural bottlenecks beyond CNN-centric models.
    \item \textbf{Actionable architectural insights.} We translate these archetypes into concrete design implications for XR-oriented accelerators and SoCs, highlighting when performance is limited by on-chip capacity, bandwidth efficiency, compute throughput, or—critically for irregular workloads—by synchronization, indexing, and phase-level overheads.
\end{itemize}
\vspace{-0.2em}
The remainder of this paper is organized as follows. Section~2 reviews related work on XR accelerators, XR SoC platforms, and system-level DSE. Section~3 introduces the workload suite and their algorithm properties. Section~4 presents our analytical modeling and mapping methodology. Section~5 reports DSE results and cross-workload energy--latency trends. Section~6 presents detailed profiling analyses on CPU and GPU platforms. Finally, Section~7 summarizes the derived workload archetypes and distills their architectural implications.

\section{Related Work}
Prior work on XR acceleration can be broadly categorized into two directions: (1) \textbf{hardware designs}, spanning task-specific accelerators and extensible SoC platforms, and (2) \textbf{system-level methodologies}, including DSE, benchmarking, and memory-centric architectures. In this section, we review representative works along both directions and highlight their methodological gaps---particularly in workload diversity, mapping methodology, and multi-granularity evidence---before positioning our workload-driven, mapping-aware characterization.

\subsection{XR Accelerators and SoC Platforms}
Most state-of-the-art XR application-oriented hardware efforts remain \emph{task- or model-specific}, optimizing for a particular kernel family such as optical flow, depth/geometry estimation, SLAM, video processing, or neural 3D reconstruction. Across these domains, bandwidth and working-set residency frequently dominate the energy--latency envelope, driving designs toward structured dataflows, aggressive on-chip buffering, operator fusion, and quantization to amortize off-chip traffic. Representative examples include pipelined cost-volume accelerators for optical flow and depth estimation on FPGA/ASIC substrates~\cite{OpticalFlow-FPGA,DepthComp-SoC}, as well as NeRF-oriented designs that co-optimize sampling/encoding, network inference, and volume rendering to reduce intermediate tensors and off-chip round trips~\cite{Moth,MetaVRain}. Beyond fixed-function pipelines, operator-level co-design studies further show that reshaping computation order and dataflow can unlock sizable efficiency gains for recurring structural motifs such as cost volumes, attention, and ray marching~\cite{CV-CIM,ViTCoD,Gen-NeRF}. Taken together, these studies demonstrate strong efficiency within their targeted kernels, but their mapping insights and provisioning strategies are often difficult to transfer across heterogeneous XR workloads that combine diverse operator types and execution patterns.

To broaden support beyond a single operator family, recent platform-level XR SoCs and accelerator prototypes integrate heterogeneous compute engines under a shared memory hierarchy~\cite{CodecAvatar-7nm,XR-NPE,SpecMeetsFlex,Aspen}. These systems combine ISP/DSP front-ends, NPUs, and SIMD/vector paths to balance domain-specific efficiency with platform scalability, with some exploring mixed-precision or posit-based execution to mitigate bandwidth pressure while maintaining accuracy across multiple perception front-end tasks~\cite{Aspen}. However, such platforms are typically demonstrated on a limited set of representative workloads and provide relatively limited coverage and mapping discussion for emerging XR workloads dominated by neural rendering and 3D reconstruction/geometry pipelines. This motivates a workload-driven characterization and mapping-aware perspective that can generalize across diverse XR operator families.

\subsection{System-Level DSE, Benchmarking, and Memory-Centric Directions}
Beyond accelerator microarchitecture, a growing line of work studies XR optimization from a system-level perspective, using design space exploration (DSE) and hardware--algorithm co-design to reason about energy--latency trade-offs under tight on-chip memory budgets. Representative efforts include SoC-level compute-in-memory benchmarking~\cite{CIM-SoC-Bench} to quantify end-to-end trade-offs under XR-oriented workloads, heterogeneous NPU--CIM designs coupled with neural architecture search~\cite{NPU-CIM} to partition models under constrained on-chip resources, and near-sensor inference that offloads early feature extraction to the sensor node to reduce system energy~\cite{NearSensor-ARVR}. While these studies provide valuable insight into partition points, cache sizing, and model adaptation, they often focus on a small set of canonical models and/or assume specific architectural substrates and partition templates, making it challenging to generalize conclusions across the broader and rapidly evolving XR workload spectrum.

Complementing these methodology-driven studies, recent XR architecture research increasingly explores memory-centric solutions. Most notably, three-dimensional (3D) stacked memory/logic integration~\cite{3D}---to directly address bandwidth and data-residency bottlenecks. By explicitly modeling vertical bandwidth and memory proximity, these studies demonstrate that moving computation closer to memory can substantially reduce off-chip traffic and yield significant speedups and energy savings compared to traditional 2D architectures. However, such benefits are typically validated on a limited set of workloads and often rely primarily on modeling-level evidence, with less multi-granularity hardware profiling and limited discussion of mapping methodologies across heterogeneous XR workloads.

In parallel, system-level benchmarking and runtime studies~\cite{XRBench,XRgo} provide end-to-end characterization under realistic latency constraints, exposing bottlenecks that may be obscured by isolated kernel analyses. Yet, these efforts are primarily scenario- or application-driven and typically offer limited architecture-facing guidance on operator-family behavior, memory-hierarchy sensitivity, or mapping-policy effects. This gap highlights the need for architecture-facing characterization that connects system-level constraints to operator- and mapping-level behavior.

In summary, prior work advances XR optimization through both hardware designs (from task-specific accelerators to extensible XR SoCs) and system-level methodologies (DSE/co-design, benchmarking, and memory-centric directions). However, many studies either focus on narrow workload subsets or lack a unified, mapping-aware methodology that generalizes across heterogeneous XR workloads. This gap motivates our workload-driven architectural classification, which connects workload diversity, multi-granularity behavior, and memory-centric trade-offs through consistent profiling and analytical exploration.

\section{Application Domain Characteristics}
Modern XR workloads span a diverse set of visual tasks, including depth estimation, optical flow/stereo matching, SLAM, and neural rendering. Many XR models contain substantial dense tensor computation, yet they also rely heavily on irregular primitives and iterative geometric routines that differ fundamentally from conventional CNN-centric inference. Representative examples include cost-volume construction, warping/resampling, non-local attention and matching, volumetric ray marching, and point-cloud alignment. These operators introduce heterogeneous locality and working-set behaviors, imposing distinct pressures on the memory hierarchy beyond standard sliding-window reuse.

We evaluate a suite of 12 representative XR workloads selected to (1) cover key stages commonly appearing in practical XR software stacks, (2) span diverse structural patterns (CNN encoder--decoders, transformer/GNN-based matchers, MLP-based neural rendering, and hybrid pipelines), and (3) reflect practical edge deployment relevance. Table~\ref{tab:workloads} summarizes the suite and reports per-workload scale metrics under a consistent batch=1 setting. Here, \emph{Wgt} denotes model parameters (or persistent state, when applicable), \emph{Act} denotes per-frame activation footprint under our execution setting, and \emph{GFLOPs} denotes per-frame floating-point operation count. For non-DNN components such as Cupoch ICP, these tensor-centric metrics are not directly comparable and are therefore omitted.
\begin{table}[t]
\centering
\small
\renewcommand{\arraystretch}{0.88}
\setlength{\tabcolsep}{5pt}

\caption{Overview of the Workload Suite and Structural Characteristics}
\label{tab:workloads}

\begin{tabularx}{\linewidth}{
  l
  >{\centering\arraybackslash}X
  c c c
}
\toprule
\textbf{Network} & \textbf{Network Description} & \textbf{Wgt (MB)} & \textbf{Act (MB)} & \textbf{GFLOPs} \\
\midrule

\multicolumn{5}{l}{\textit{Depth Estimation}} \\
Monodepth2 & ResNet encoder--decoder (U-Net style) & 53.87 & 69.42 & 11.73 \\
HR-Depth   & Encoder--decoder CNN (multi-scale)    & 53.64 & 409.81  & 22.43 \\

\addlinespace[0.3em]
\multicolumn{5}{l}{\textit{Optical Flow and Stereo Matching}} \\
PWC-Net     & Pyramid CNN + cost volume             & 35.76 & 1833.46 & 169.48 \\
RAFT-Stereo & Recurrent CNN + 3D correlation volume & 42.27 & 618.05  & 84.44  \\

\addlinespace[0.3em]
\multicolumn{5}{l}{\textit{SLAM Frontend: Feature Matching and Tracking}} \\
LightGlue & CNN backbone + transformer matcher  & 12.21 & 954.02  & 90.44  \\
SuperGlue & Graph neural network matcher        & 11.82 & 996.33 & 104.33 \\
LoFTR     & Detector-free transformer matcher   & 87.98 & 5105.88 & 427.96 \\

\addlinespace[0.3em]
\multicolumn{5}{l}{\textit{SLAM Backend and Mapping}} \\
TartanVO   & PWC-Net matcher + ResNet50 pose head & 19.96 & 1177.81 & 30.94 \\
Cupoch ICP & CUDA-based point-cloud ICP pipeline  & --    & --      & --    \\

\addlinespace[0.3em]
\multicolumn{5}{l}{\textit{Neural Rendering (NeRF-style MLPs)}} \\
NeRF(and TinyNeRF)     & Fully-connected MLP + ray marching       & 1239.42 & 11318 & 813.54 \\

\addlinespace[0.3em]
\multicolumn{5}{l}{\textit{Vision Backbones}} \\
ViT & Vision Transformer encoder (classifier) & 25.63 & 1486.87 & 18.20 \\

\bottomrule
\end{tabularx}
\end{table}
From an architectural perspective, rather than grouping workloads solely by algorithmic function, we organize them into structural families that exhibit distinct dataflow patterns and working-set characteristics.

\textbf{Encoder--decoder CNNs.}
Models such as Monodepth2~\cite{Monodepth2} and HR-Depth~\cite{HR-Depth} adopt U-Net-style architectures with multi-scale feature pyramids and skip connections. Compared to classification backbones (e.g., ResNet), the decoder path maintains high-resolution intermediate feature maps, resulting in a substantially larger activation working set. Consequently, performance and energy are often sensitive to on-chip capacity and the ability to keep key intermediates resident to reduce off-chip traffic.

\textbf{Cost-volume and warping-based models.}
Optical flow and stereo workloads such as PWC-Net~\cite{PWCNET}, RAFT/RAFT-Stereo~\cite{raft}, and hybrid pipelines such as TartanVO~\cite{TartanVO} explicitly construct correlation/cost volumes and use warping or resampling for feature alignment. Cost-volume construction involves gather-heavy access over a search window, while warping performs coordinate-based sampling. These steps disrupt the regular spatial locality of standard convolutions and can limit reuse realization under conventional mapping strategies.

\textbf{Feature matching networks (transformer/GNN-based).}
LightGlue~\cite{Lightglue}, LoFTR~\cite{LoFTR}, and SuperGlue-style matchers~\cite{SuperGlue} combine CNN features with transformer- or graph-based matching modules, while ViT~\cite{ViT} represents a pure token-based backbone. Operators such as patchify, self-/cross-attention, and graph message passing require maintaining multiple intermediate tensors (e.g., query/key/value or node/edge states) and often induce non-local access patterns. As a result, their bottlenecks are frequently shaped by working-set residency and memory movement rather than convolution-like reuse.

\textbf{Neural volumetric rendering (NeRF-style).}
NeRF~\cite{Nerf} and TinyNeRF~\cite{TinyNeRF} represent neural graphics workloads composed of ray-marching loops coupled with small MLP evaluations. While the MLP itself is typically compact, the overall execution is dominated by massive numbers of per-ray sample evaluations and intermediate accumulation, which can generate a ray-dominant activation stream with phase-dependent compute and memory behaviors. This structure can decouple compute intensity from weight bandwidth and makes performance sensitive to sampling strategy and on-chip buffering of per-ray states.

\textbf{Iterative geometric solvers.}
Workloads such as Cupoch ICP~\cite{CupochPyPI} rely on iterative optimization (e.g., Iterative Closest Point), where each iteration performs nearest-neighbor search, rigid transform estimation, and error evaluation. Architecturally, these pipelines exhibit complex control flow, sparse point-cloud operations, and small-scale linear algebra, and they are often latency-bound rather than throughput-bound. Their performance is therefore shaped by irregular memory access, synchronization, and branch/control overheads in addition to raw compute capability.

\section{Architectural Modeling and High-Level Parameterized DSE Framework}
Evaluating XR workloads at the architectural level requires characterizing how each application stresses both the compute engine and the memory hierarchy—most notably off-chip bandwidth demand, on-chip capacity pressure, and compute throughput requirements. XR workloads are structurally heterogeneous: their operators and graph topologies are highly diverse, and the dominant bottleneck often shifts across stages (e.g., from compute-limited to bandwidth- or capacity-limited). This diversity makes it difficult to compare applications or extract generalizable trends using ad-hoc, operator-specific analysis.
To enable systematic and comparable cross-workload evaluation, we develop a lightweight, C++-based DSE framework—a fast analytical design-space exploration model that avoids RTL or cycle-accurate simulation while capturing the key energy–latency trade-offs under tight on-chip memory budgets. The framework is designed to cover a broad set of vectorizable operator patterns commonly found in XR workloads, enabling consistent cost accounting across heterogeneous graphs. Our objective is not to claim cycle-level prediction accuracy; rather, we provide an interpretable, hierarchy-aware cost model that supports rapid sweeping over a large design space and exposes each workload’s structural sensitivity to on-chip capacity, memory bandwidth, and mapping choices. This capability is essential for identifying bottleneck tipping points, explaining diminishing returns, and producing comparable metrics across diverse XR networks.
In the rest of this section, we present the design rationale and key components of the proposed DSE model in detail.

\begin{figure}[hbtp]
    \centering
    \includegraphics[width=\linewidth]{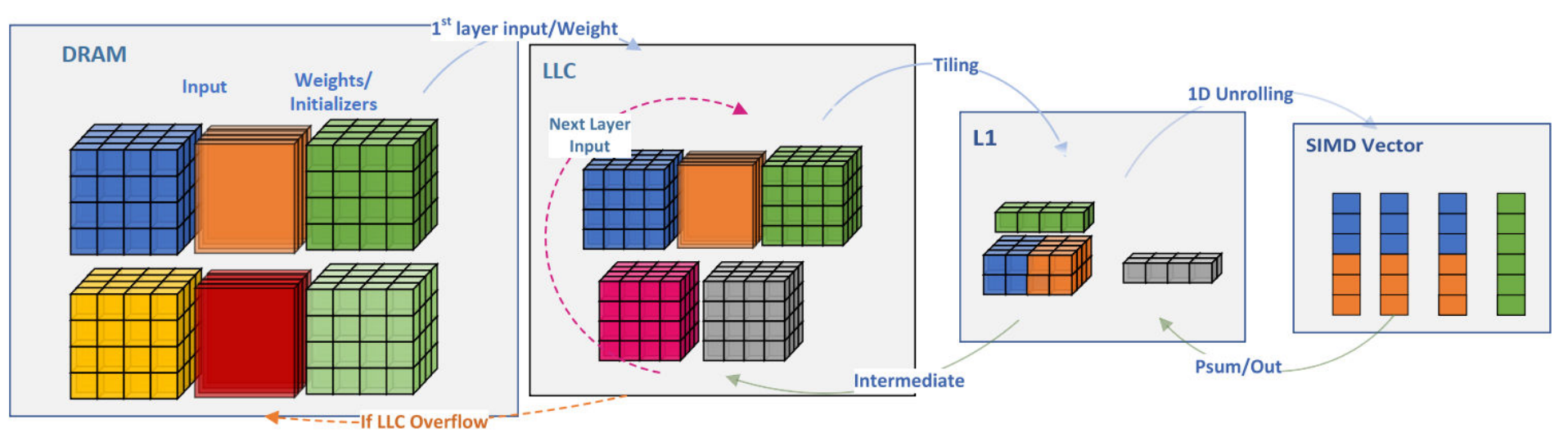}
    \vspace{-1em}
    \caption{Near-memory SIMD Modeling Template Dataflow Overview}
    \label{fig:placeholder}
\end{figure}

\subsection{Architecture Template Overview}
To instantiate DSE modeling on a consistent execution substrate, we adopt a single-core SIMD vector processor as the baseline architectural template (inspired by low-power vector architectures\cite{cavalcante2019ara}\cite{Soft-SIMD}\cite{Block-Scaled}), coupled with a configurable multi-level on-chip memory hierarchy.
We choose a SIMD-style template primarily for modeling generality and mapping consistency across XR operator classes. XR pipelines are not solely composed of dense MAC-centric kernels; they also include many vectorizable primitives that are awkward to express under a rigid systolic-array-style dataflow. In practice, supporting such primitives on a MAC-only TPU often requires additional DSP/SIMD side units and operator-specific dataflow tailoring across stages, which would complicate the mapping policy and blur this paper’s goal of using a unified evaluation model to expose architectural sensitivities under controlled on-chip budgets. Importantly, this template serves as an analysis vehicle for Fast-DSE rather than a claim of a uniquely optimal XR accelerator design.

\textbf{SIMD Vector Processor Block:}
The compute core is a wide SIMD vector processor with a configurable number of lanes (default: 16). The vector register file (VRF) is 512-bit wide and supports multiple data types (e.g., FP32/FP16/INT8). The core is assumed to provide common XR-relevant vector primitives, including fused multiply–add, elementwise arithmetic, comparisons/min–max, reductions, and data-movement operations (e.g., permute and gather/scatter-style accesses). This design accommodates both compute-dense Conv/GEMM phases and transform-/resampling-/matching-heavy phases that frequently appear in XR networks, while keeping the execution model uniform across operator families.

\textbf{L1 On-Chip Memory:}
The L1 on-chip memory acts as a software-managed scratchpad, explicitly controlled by the mapper, with a configurable capacity. L1 primarily serves as the inner-tile working buffer, enabling fine-grained data movement, partitioning, and flexible tensor tiling for large operators. It improves locality for hot tensor slices and short-lived intermediates, and it serves as a critical design parameter that determines whether an operator can exploit inner-loop reuse without spilling to lower levels.

\textbf{LLC Reuse Buffer (Software-Managed):}
The last-level on-chip storage (LLC) provides 16–64 MB of capacity and interfaces between the L1 scratchpads and DRAM. In this work, the LLC is modeled as a software-visible reuse buffer rather than a hardware-managed cache. Its primary role is to host large intermediates or weights that exceed L1 capacity, thereby providing an analytical upper bound on data reuse based on capacity residency. By treating the LLC as a capacity-limited buffer, our DSE abstracts away specific hardware replacement policies (e.g., LRU) to focus on how capacity allocation and management influence DRAM traffic and overall energy efficiency.

\textbf{DRAM Main Memory:}
DRAM is the lowest-level storage in the hierarchy, with access latency and energy consumption much higher than those of on-chip memories, but it is necessary for storing large feature maps and model parameters. We assume that all pretrained weights and network inputs are initially stored in DRAM. DRAM parameters (energy, bandwidth, and latency) are derived from Ramulator\cite{ramulator} and DDR4-standard DRAMPower\cite{DRAMPowerTool} datasheets to ensure consistency with mainstream technology assumptions.

\subsection{Dataflow and Mapping Strategy}

To ensure cross-workload comparability, our framework adopts a uniform and conservative mapping policy. Rather than applying network-specific scheduling or layout optimizations, we restrict mapping choices to a small, shared rule set (stationary selection, L1 tiling, and limited tile-local fusion). This design makes the subsequent $(L1, LLC)$ sweeps primarily reflect workload-intrinsic sensitivity to on-chip capacity and off-chip traffic, instead of per-model hand tuning.
The modeling flow begins with an intermediate representation (IR) of each network. For every operator, we parse the operator type, tensor shapes, and constant initializers (e.g., weights and lookup tables), and construct a producer–consumer dependency DAG. We then fix a deterministic execution order using a topological traversal, which serves as the common schedule for tensor-liveness analysis and hierarchy-level traffic accounting. 

Our baseline hierarchy is DRAM $\rightarrow$ LLC $\rightarrow$ L1 $\rightarrow$ SIMD, similar in structure to the explicit memory-hierarchy abstractions used in prior mapping frameworks (e.g.,~\cite{timeloop}) for reasoning about data movement and locality. Unlike hardware caches with implicit replacement, we treat the LLC as a software-managed, capacity-bounded reuse buffer for controlled residency analysis. Concretely, we approximate ideal retention under a capacity constraint using tensor liveness along the fixed execution order: a tensor becomes live when produced and is released after its last consumer, and tensors spill to DRAM when the live footprint exceeds the LLC capacity. This abstraction removes conflict/replacement dynamics, providing a stable, capacity-driven view for analyzing capacity bottlenecks.

To keep the model lightweight for large parameter sweeps, we use a linearized energy accounting scheme. Memory energy is computed from transferred bytes, using separate read/write coefficients drawn from CACTI-like normalized SRAM/DRAM PPA tables. Compute energy is estimated using normalized per-operation energy costs drawn from the same technology/PPA library (e.g., MAC- or SIMD-op-equivalent energy), with coefficients chosen based on the compute template and operation class. 

Fig. 2 summarizes our uniform mapping optimization flow, which consists of three components: L1 tiling, stationary selection, and limited kernel fusion. We apply the same tiling-and-stationary procedure to all workloads to preserve cross-network comparability. As shown in Fig. 2, the L1–SIMD interaction determines whether a candidate tile can fit within the effective L1 budget; the detailed tiling search is described in Section 4.3. Here we describe the two lightweight mapping rules that accompany tiling and are applied consistently across all workloads.

\textbf{Footprint-based stationary selection}
For Conv/GEMM-like operators, we select between output-stationary (OS) and weight-stationary (WS) using a simple weight-footprint heuristic. The key rationale is that the stationary choice should not compromise the feasibility and quality of subsequent L1 tiling: L1 must simultaneously accommodate at least one input tile, one output/PSUM tile, and (if WS is used) a weight tile. Therefore, we use the layer’s weight footprint $B_w$ as a conservative indicator of whether weights can be kept resident in L1 without crowding out activation tiles.
Concretely, we choose WS when $B_w < \rho \cdot L1_{\mathrm{eff}}$ (with a fixed $\rho=0.5$ across all experiments), which reserves the remaining L1 capacity for input/output tiles and bookkeeping. Under this condition, keeping weights stationary maximizes weight reuse across the inner compute loops and reduces repeated L1$\leftrightarrow$LLC transfers of weights. Otherwise, when $Bw$ is large relative to L1, we choose OS so that L1 capacity is prioritized for output/PSUM accumulation and activation tiles, avoiding pathological tilings that either become infeasible or require excessive tile fragmentation. In our framework, this stationary decision affects only reuse assumptions and hierarchy-level traffic accounting, while the concrete tile feasibility and tile shapes are determined by the unified L1 tiling procedure in Section~4.3.

\textbf{Legality-preserving local fusion.}
To reduce avoidable L1 and LLC round trips in common XR operator chains, we apply a small set of local fusion rules. Fusion is restricted to simple, tile-local producer–consumer patterns where the consumer can be evaluated immediately on the producer’s tile without changing tensor shape or introducing additional large live buffers. Concretely, we fuse (i) Conv/GEMM followed by a unary activation, and (ii) elementwise add followed by activation, whenever the two operators share the same tile shape and the fused execution does not increase the number of live tensors or require materializing extra multi-input intermediates within a tile. We explicitly avoid fusion across reshape, concatenation, or other layout/materialization boundaries, since such operators typically break tile alignment and complicate residency accounting.

\begin{figure}[htbp]
  \centering
  \newcommand{\FigH}{5cm} 

  \begin{minipage}[t]{0.65\linewidth}
    \vspace{0pt}\centering
    \includegraphics[height=\FigH]{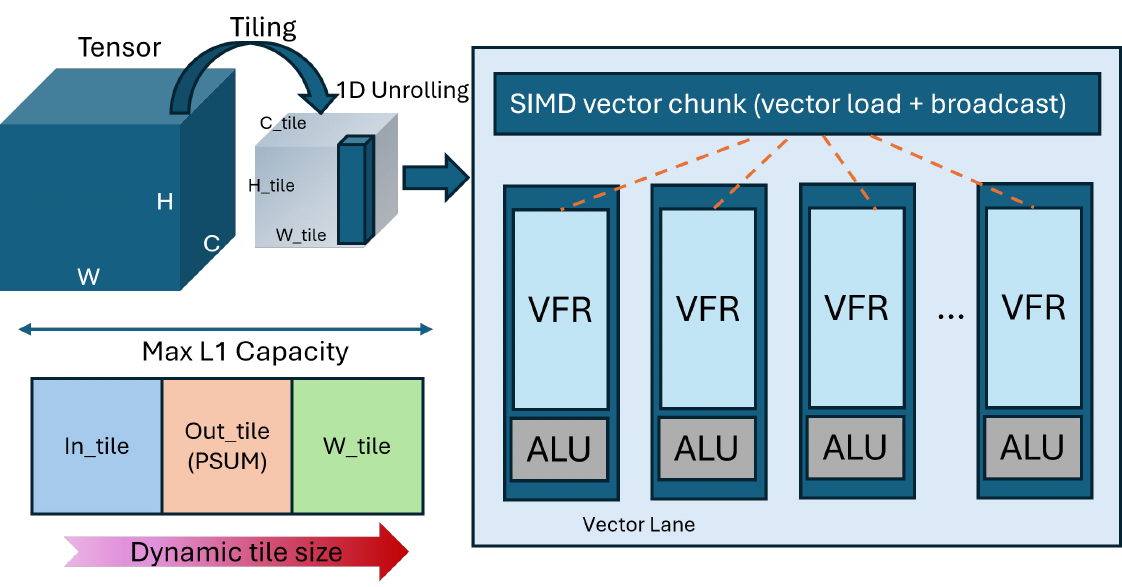}
  \end{minipage}\hfill
  \begin{minipage}[t]{0.35\linewidth}
    \vspace{0pt}\centering
    \includegraphics[height=\FigH]{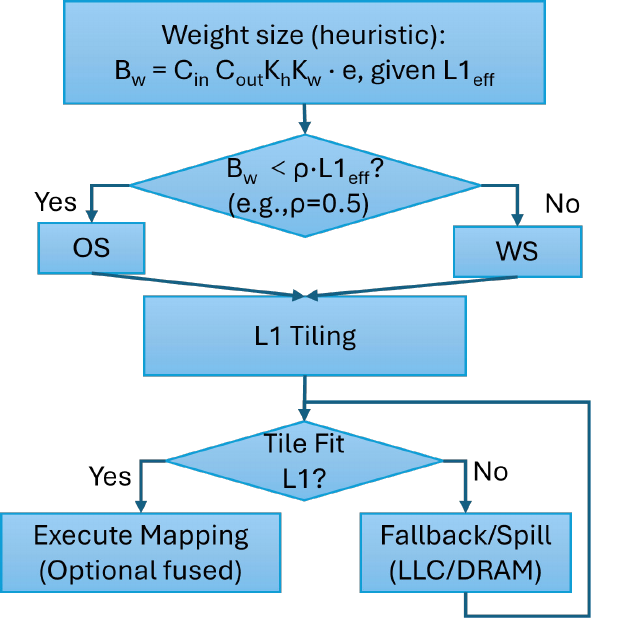}
  \end{minipage}

  \caption{L1-resident tiling and stationary execution flow on the SIMD engine(Conv layer case).}
  \label{fig:tiling-stationary}
\end{figure}

\subsection{Tiling Automation}
We further detail the L1 tiling automation used in our mapper. To preserve cross-network comparability, we adopt a uniform and conservative mapping policy; under this policy, large tensor MAC operators (e.g., Conv and GEMM/MatMul) often cannot be directly mapped because the combined working set of inputs, weights, and outputs may exceed the effective L1 capacity. We therefore introduce an automated L1 tiling mechanism that, given an L1 capacity budget, (i) guarantees mapping legality and (ii) searches for a high-quality tiling configuration using a workload-agnostic procedure. Due to space limitations, we describe convolution tiling as a representative case; the same mechanism extends to GEMM/MatMul with analogous blocking variables and the same feasibility test. Fig. 3(a) summarizes the tiling variables and the L1 feasibility constraint, while Fig. 3(b) (Algorithm 1) illustrates the search flow.

For a Conv tile to execute within our L1–SIMD template, L1 must simultaneously accommodate three footprints: (i) the input activation tile, (ii) the corresponding weight tile, and (iii) the output/partial-sum (PSUM) tile. As summarized in Fig. 3(a), we parameterize a candidate tiling by blocking along channel and output-spatial dimensions. Importantly, the required input-tile footprint is not an independent choice: it is induced by the output tile through the convolution receptive field (stride/kernel/padding), and therefore captures halo/boundary expansion. A tiling is legal only if the combined footprint of these three tiles fits within the effective L1 budget (Fig. 3(a)); otherwise the mapping would require spilling and cannot be executed as an L1-resident tile.

Among legal tilings, we rank candidates using a lightweight surrogate objective:
\vspace{-0.5em}
\begin{equation}
J(t) = \alpha \cdot \mathrm{Compute}(t) + \beta \cdot \mathrm{Bytes}(t) + \gamma \cdot N_{\mathrm{tiles}}(t).
\end{equation}

Here, Compute(t) and Bytes(t) reuse the hierarchy-level accounting described in Section 4.2, serving as proxies for arithmetic work and L1-boundary traffic under the chosen blocking. The construction of this objective follows the design philosophy of scheduling frameworks such as the Halide auto-scheduler \cite{HalideLectureCS348V, Halide}, which combines estimates of computation and memory access to guide scheduling and mapping decisions, and has been widely adopted in tensor program schedulers and accelerator mapping frameworks. The N\_tiles(t) term is derived from the number of tiles implied by the blocking and regularizes against overly fragmented tilings, which tend to increase per-tile control/setup overhead and amplify boundary/halo redundancy as discussed in prior mapping studies\cite{timeloop}. The weights $(\alpha,\beta,\gamma)$ are fixed globally across all workloads and capacity points; we scale them once so that the three terms are of comparable magnitude for a representative feasible tiling, and we do not tune them per workload.

\begin{figure}[t]
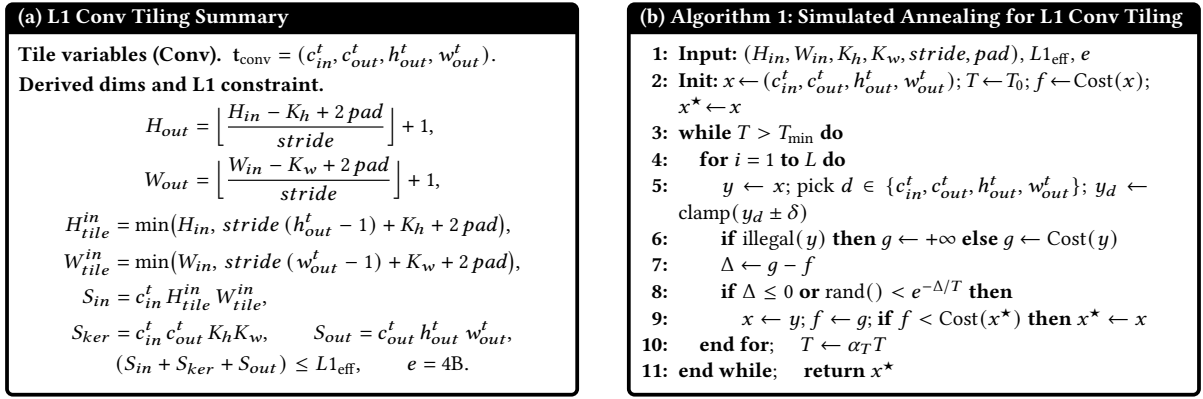

  \centering

  \begin{minipage}[t]{0.48\linewidth}
    \vspace{0pt}
    \begin{tcolorbox}[
      enhanced,
      equal height group=tilingpair,
      height fixed for=first,
      colback=white,colframe=black,
      boxsep=0.8pt, arc=1.5pt,
      left=3pt,right=3pt,top=3pt,bottom=3pt,
      title={\footnotesize (a) L1 Conv Tiling Summary},
      fonttitle=\bfseries\footnotesize,
      colbacktitle=black, coltitle=white
    ]
      \footnotesize
      \textbf{Tile variables (Conv).}\;
      $\mathbf t_{\mathrm{conv}}=(c_{in}^t,c_{out}^t,h_{out}^t,w_{out}^t).$

      \vspace{0.4ex}
      \textbf{Derived dims and L1 constraint.}
      \[
      \begin{aligned}
      H_{out} &= \Big\lfloor\frac{H_{in}-K_h+2\,pad}{stride}\Big\rfloor + 1, \\
      W_{out} &= \Big\lfloor\frac{W_{in}-K_w+2\,pad}{stride}\Big\rfloor + 1,
      \end{aligned}
      \]
      \[
      \begin{aligned}
      H^{in}_{tile} &= \min\!\big(H_{in},\, stride\,(h_{out}^t-1)+K_h+2\,pad\big),\\
      W^{in}_{tile} &= \min\!\big(W_{in},\, stride\,(w_{out}^t-1)+K_w+2\,pad\big),
      \end{aligned}
      \]
      \[
      \begin{aligned}
      S_{in}  &= c_{in}^t\,H^{in}_{tile}\,W^{in}_{tile}, \\
      S_{ker} &= c_{in}^t\,c_{out}^t\,K_hK_w, \qquad
      S_{out} = c_{out}^t\,h_{out}^t\,w_{out}^t,
      \end{aligned}
      \]
      \[
      (S_{in}+S_{ker}+S_{out}) \le L1_{\mathrm{eff}},\qquad e=4\mathrm{B}.
      \]
    \end{tcolorbox}
  \end{minipage}\hfill
  \begin{minipage}[t]{0.48\linewidth}
    \vspace{0pt}
    \begin{tcolorbox}[
      enhanced,
      equal height group=tilingpair,
      height fixed for=first,
      colback=white,colframe=black,
      boxsep=0.8pt, arc=1.5pt,
      left=3pt,right=3pt,top=2pt,bottom=2pt,
      title={\footnotesize (b) Algorithm 1: Simulated Annealing for L1 Conv Tiling},
      fonttitle=\bfseries\footnotesize,
      colbacktitle=black, coltitle=white
    ]
      \footnotesize
      \renewcommand{\arraystretch}{1.00} 
      \setlength{\tabcolsep}{2pt}       

      \begin{tabular}{@{}r p{0.86\linewidth}@{}}
      \textbf{1:} & \textbf{Input:} $(H_{in},W_{in},K_h,K_w,stride,pad)$, $L1_{\mathrm{eff}}$, $e$ \\
      \textbf{2:} & \textbf{Init:} $x\!\leftarrow\!(c_{in}^t,c_{out}^t,h_{out}^t,w_{out}^t)$; $T\!\leftarrow\!T_0$; $f\!\leftarrow\!\mathrm{Cost}(x)$; $x^\star\!\leftarrow\!x$ \\
      \textbf{3:} & \textbf{while} $T>T_{\min}$ \textbf{do} \\
      \textbf{4:} & \quad \textbf{for} $i=1$ \textbf{to} $L$ \textbf{do} \\
      \textbf{5:} & \quad\quad $y\leftarrow x$; pick $d\in\{c_{in}^t,c_{out}^t,h_{out}^t,w_{out}^t\}$; $y_d\leftarrow \mathrm{clamp}(y_d\pm\delta)$ \\
      \textbf{6:} & \quad\quad \textbf{if} illegal$(y)$ \textbf{then} $g\leftarrow+\infty$ \textbf{else} $g\leftarrow\mathrm{Cost}(y)$ \\
      \textbf{7:} & \quad\quad $\Delta\leftarrow g-f$ \\
      \textbf{8:} & \quad\quad \textbf{if} $\Delta\le 0$ \textbf{or} $\mathrm{rand()}<e^{-\Delta/T}$ \textbf{then} \\
      \textbf{9:} & \quad\quad\quad $x\leftarrow y$; $f\leftarrow g$; \textbf{if} $f<\mathrm{Cost}(x^\star)$ \textbf{then} $x^\star\leftarrow x$ \\
      \textbf{10:} & \quad \textbf{end for}; \quad $T\leftarrow \alpha_T T$ \\
      \textbf{11:} & \textbf{end while}; \quad \textbf{return} $x^\star$ \\
      \end{tabular}
    \end{tcolorbox}
  \end{minipage}

  \caption{Automatic L1 tiling for convolution layers.
(a) Conv tile variables and the L1 capacity constraint from input/weight/output footprints.
(b) Simulated annealing search that explores feasible tiles under the constraint and returns a low-cost tiling. All SA hyperparameters ($T_0$, $T_{\min}$, $\alpha_T$, $L$, and $\delta$) are fixed globally across all workloads.}
  \label{fig:tiling-sa}
  \vspace{-0.8em}
\end{figure}

We employ simulated annealing (SA) as a lightweight meta-heuristic to search the discrete, capacity-constrained tiling space under a fixed evaluation budget, following common practice in compiler- and auto-tuning--driven design space exploration (e.g., TVM-style scheduling search\cite{TVM}). Algorithm~1 (Fig.~3(b)) summarizes the tiling search procedure used in our mapper.
For a given layer and $L1_{\mathrm{eff}}$, the tiling state is
$x=(c^{t}_{\mathrm{in}}, c^{t}_{\mathrm{out}}, h^{t}_{\mathrm{out}}, w^{t}_{\mathrm{out}})$.
We initialize SA from a trivially feasible state by shrinking blocking factors until the L1 feasibility constraint in Fig.~3(a) is satisfied.
SA then explores neighboring tilings by perturbing one blocking dimension per move; candidates that violate the L1 constraint are treated as illegal (infinite cost), while feasible candidates are scored by $J(\cdot)$ and accepted according to the standard SA criterion under a fixed cooling schedule.
The best feasible tiling $x^*$ found within the search budget is returned.

For each Conv/GEMM-like operator and each L1 capacity point in the sweep, the tiler returns the best feasible blocking found under this fixed SA budget.
The selected blocking is then passed to the model for consistent traffic and energy accounting across workloads.

\section{DSE result analysis}
In this section, we use the fast analytical model from Section~4 to study how on-chip memory capacity shapes the energy and memory-side latency behavior of XR workloads. Our on-chip SRAM energy/area assumptions are anchored to publicly reported technology data and scaling trends~\cite{cactipp,cfet,l1}. For each network, we sweep the L1 scratchpad capacity from 16\,KB to 256\,KB and the LLC capacity from 16\,MB to 64\,MB, spanning configurations from a small baseline up to the largest capacities considered feasible under this technology/area baseline.\footnote{In a deployment-oriented setting, different networks would naturally be provisioned with different on-chip memory configurations. For example, workloads with large data volume or large intermediate footprints may require a larger memory configuration, while smaller-footprint workloads could operate efficiently with a smaller configuration. Here we intentionally use a uniform capacity grid to enable a controlled sensitivity study and cross-workload comparison, rather than per-network tuning.}

Throughout this section, DRAM capacity is fixed at 8 GB. Under batch-1 inference, this capacity is sufficient to accommodate the working sets of all evaluated workloads, so our results avoid secondary effects caused purely by insufficient DRAM capacity, such as footprint overflow. Consequently, changes in DRAM energy and memory-side latency primarily reflect differences in off-chip traffic volume and access behavior, rather than address-space capacity effects. This setup enables a controlled, cross-application comparison of hierarchy-level behavior under a uniform capacity assumption.

For total energy, we report the sum of L1 energy, LLC energy, DRAM energy, and vector-core energy as defined in Section 4. For latency, since the analytical model is not cycle-accurate across heterogeneous clock domains, we avoid interpreting on-chip cycle counts as absolute time. Instead, we adopt a roofline-style latency proxy and define memory-side latency as the maximum of on-chip service time and DRAM service time, computed from the corresponding traffic volumes and service bandwidths. This proxy is used only as a relative indicator to compare design points within each workload. It captures the practical reality that, once off-chip traffic is non-trivial, DRAM service time can dominate the effective memory-side stall component.

\subsection*{5.1\quad Memory-side energy scaling with L1/LLC capacity}
Fig. 4 summarizes the energy impact of the capacity sweep for representative workloads in our XR suite.\footnote{Our DSE focuses on kernels that can be consistently represented under our tiling-based SIMD/tensor abstraction to enable controlled cross-workload comparison. The point-cloud ICP pipeline is excluded from DSE because its sparse, control-flow-heavy kernels (e.g., nearest-neighbor search) are not captured by this abstraction; we still analyze it via hardware profiling in Section 6. TinyNeRF is omitted because it follows the same NeRF-style kernel structure as NeRF and does not introduce a distinct capacity-sensitivity regime.}
Figure 4(a) reports the energy breakdown at the baseline configuration (32 KB L1 and 16 MB LLC) and at the most energy-efficient configuration for each application. This comparison highlights which component dominates memory energy at baseline and which components are reduced when moving toward the best design point. Figure 4(b) visualizes the two-dimensional energy landscape over the evaluated L1 and LLC grid as a residency-style heatmap, with total energy encoded by color (darker indicates higher energy). Each cell reports normalized total energy under a specific provisioning point. In this view, sharp transitions between neighboring cells indicate a regime change in working-set residency, where a subset of recurrent intermediates becomes on-chip resident, while relatively uniform regions indicate diminishing marginal returns under further scaling.

\begin{figure}[htbp]
    \centering
    \begin{subfigure}{\linewidth}
        \centering
        \includegraphics[width=\linewidth]{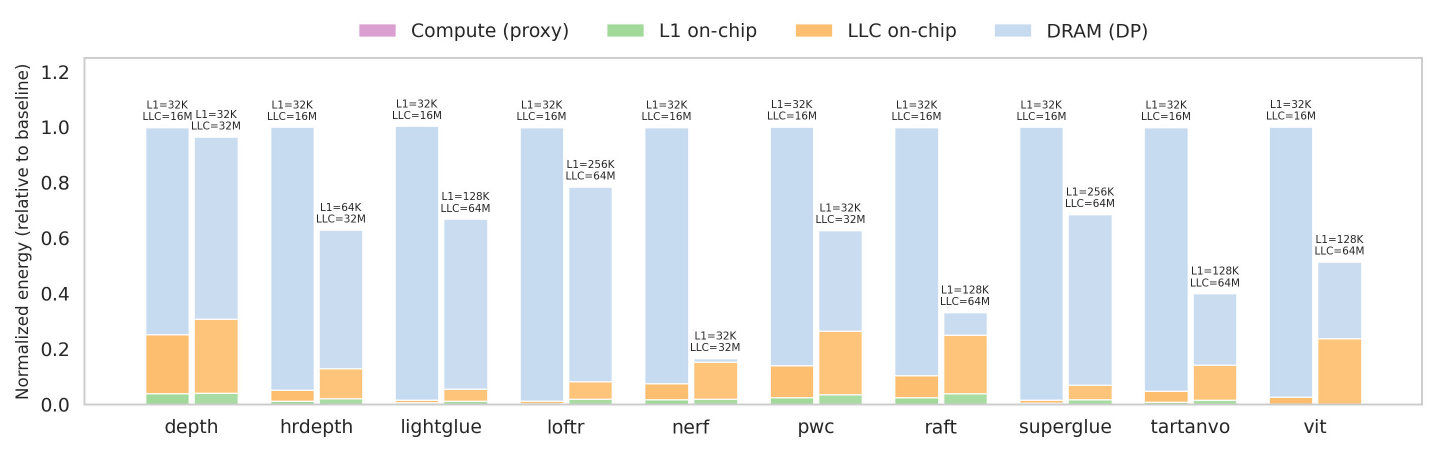}
        \caption{Energy breakdown (global)}
        \label{fig:4a}
    \end{subfigure}

    \vspace{4pt}

    \begin{subfigure}{\linewidth}
        \centering
        \includegraphics[width=\linewidth]{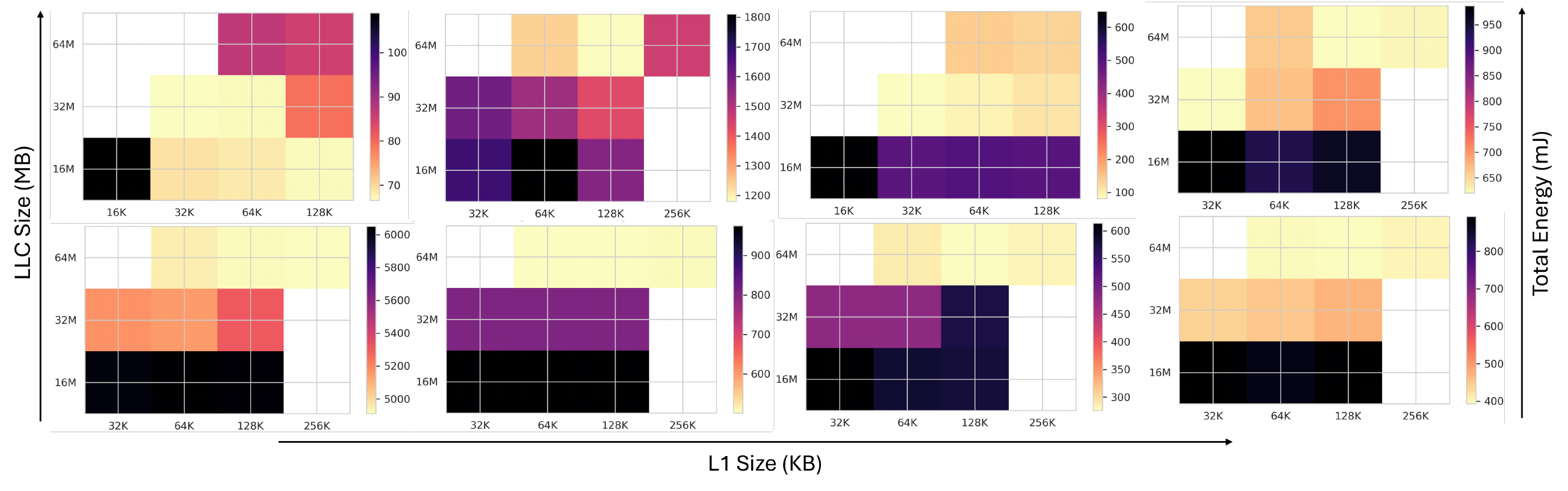}
        \caption{2D Energy Heatmap for L1 vs LLC(from top-left to bottom-right: monodepth2, lightglue, Nerf,   pwcnet, loftr, ViT, raft-stereo, tartanvo.) }
        \label{fig:4b}
    \end{subfigure}

    \caption{Energy vs. on-chip capacity: breakdown and L1–LLC capacity across workloads.}
    \label{fig:fig4}
\end{figure}
Under our energy model, DRAM is the dominant contributor to memory energy at the baseline configuration for most workloads, followed by LLC and then L1. By comparing the baseline energy breakdown to the best-case configuration, we observe that increasing LLC capacity often shifts a substantial portion of memory energy from DRAM to LLC, since more intermediate tensors become resident on chip. For networks whose effective reuse footprint fits within tens of megabytes, this migration can reduce DRAM energy to a negligible fraction at the optimal configuration. In contrast, for large data-volume workloads, DRAM remains a significant energy component even at the maximum LLC capacity we sweep, indicating that their effective reuse footprint exceeds the available on-chip storage under our mapping policy.

We next examine sensitivity along the L1 axis in Fig. 4(b). Overall, the impact of L1 capacity is strongly conditioned on LLC size. When LLC is small (16 MB), L1 capacity has limited effect on total energy for most workloads because DRAM traffic dominates the energy budget. One notable exception is Monodepth2: its intermediate footprints are relatively smaller, so even a small LLC captures a larger fraction of reuse, making residual sensitivity to on-chip buffering more visible. Once LLC increases to 32 MB and above, L1 behavior becomes workload-specific. Some workloads, such as LoFTR and ViT, are largely insensitive to L1 capacity because their dominant reuse is mediated by LLC residency, while L1 mainly buffers short-lived tiles. Other workloads exhibit mild non-monotonic behavior, where the energy-optimal L1 size depends on LLC capacity. When LLC is small, enlarging L1 can temporarily increase total energy because larger tiles may keep partial sums and intermediate fragments resident longer, increasing on-chip access cost while DRAM traffic remains high. In addition, larger L1 capacity increases access and leakage overhead in the energy model. Once LLC becomes large enough to hold the key intermediate tensors, the same increase in L1 can reduce redundant refetches and lower total energy. This grow-to-fit effect arises from discrete tiling choices under a fixed mapping strategy.
Along the LLC dimension, the heatmap reveals a more systematic sensitivity than along L1 for most workloads: for a fixed L1, increasing LLC often moves the configuration from a high-energy region to a lower-energy region, and this trend is broadly consistent across L1 choices. A common signature is a noticeable transition band between the 16 MB and 32 MB LLC regimes, suggesting that a non-trivial subset of recurrent intermediates begins to remain on chip and avoid repeated DRAM refetches. Beyond this transition region, behaviors are more workload dependent. For some workloads (e.g., NeRF and PWC-Net), most of the convertible reuse is already captured around the 32 MB regime and further LLC increases provide diminishing returns. Others (e.g., LoFTR, ViT, RAFT-Stereo, and TartanVO) remain effectively DRAM-active around 32 MB and only enter a lower-energy region closer to the 64 MB regime, consistent with a larger aggregate footprint of concurrently live intermediates. 
Following these observations, we group workloads into three recurring capacity-response regimes  
(1) \emph{Early-saturating / diminishing-returns workloads} (e.g., Monodepth2, HRDepth and MLP-based NeRF) quickly enter a relatively uniform region in the heatmap: modest LLC provisioning captures most of the convertible reuse, and further L1/LLC scaling yields limited marginal energy reduction. 
(2) \emph{Capacity-gated, LLC-driven DRAM displacement workloads} (e.g., RAFT-Stereo, PWC-Net, ViT, and TartanVO) exhibit clear transition bands along the LLC axis: once LLC crosses a working-set boundary, a large portion of DRAM energy is displaced by LLC energy, while L1 remains weakly sensitive or only conditionally beneficial after LLC becomes sufficiently large. 
(3) \emph{Persistent DRAM-active / reuse-realization-gap workloads} (e.g., LightGlue, SuperGlue, and LoFTR) remain DRAM-dominated across the explored grid: although total energy can vary with L1/LLC due to on-chip access cost and discrete tiling effects, passive cache scaling alone does not yield a collapse of DRAM energy comparable to group (2), suggesting that reuse is difficult to materialize under the observed execution structure.

\subsection*{5.2\quad Energy--Latency Interaction and Pareto Trade-off}

Section 5.1 shows that increasing on-chip capacity (L1/LLC) can reduce DRAM traffic and thus improve energy efficiency. We now extend the question to performance: when DRAM traffic is reduced by capacity provisioning, does the memory-side latency improve accordingly? Which workloads expose energy–latency trade-offs, and do different workloads show distinct trajectory shapes?
Fig. 5 plots total energy versus memory-side latency (as defined by the roofline-style max of on-chip and DRAM service time) for each cache configuration. Both axes are normalized per workload, so the focus is not on cross-workload absolute values but on the trajectory and Pareto shape within each workload.\footnote{The remaining workloads in our suite exhibit similar qualitative trends; we therefore show only four representative networks to keep the discussion concise while preserving the key behaviors.}

\begin{figure}[htbp]
  \centering
  \includegraphics[width=\linewidth]{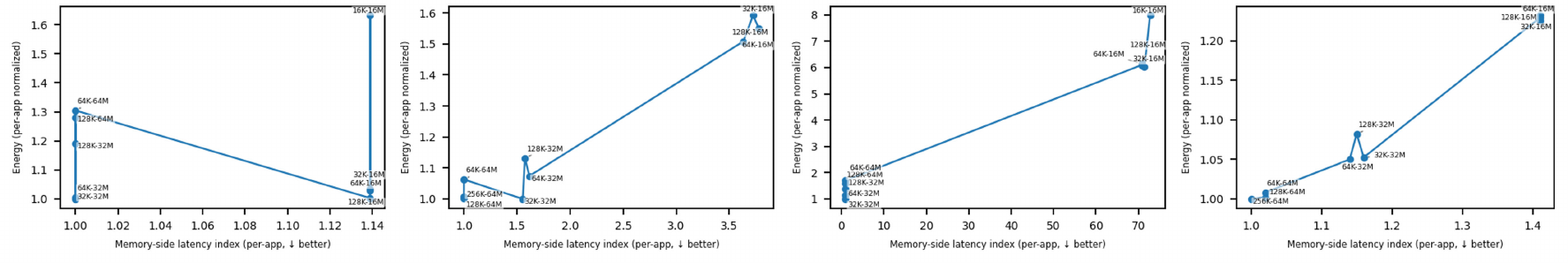}
  \caption{Pareto fronts of energy vs memory-side latency(Workload from left to right: LoFTR, NeRF, PWC-Net, and Monodepth2).}
\end{figure}

Most workloads follow a near-diagonal joint-improvement trend: configurations that reduce DRAM traffic tend to improve both energy and memory-side latency at the same time. NeRF is a clear example—when on-chip caches are small, memory-side latency is high, but with sufficiently large LLC, both energy and latency drop sharply, reflecting that cache residency can effectively suppress repeated off-chip transfers. LoFTR shows a similar, though more gradual, pattern: as LLC increases, both energy and latency proxy steadily improve, until reaching a plateau once the key intermediate features fit on chip. In these cases, the most energy-efficient configuration is close to the lowest-latency one for the explored design space.

However, this is not always the case. Fig. 5 also reveals workloads where the trade-off is more complex and requires explicit multi-objective balancing. For example, PWC-Net demonstrates a two-regime behavior: increasing LLC leads to a clear drop in latency, but the configuration with minimum energy does not always match the one with minimum latency, mainly due to how on-chip energy and tiling interact. Both points appear on the Pareto frontier, so the preferred operating point depends on whether energy or latency is the system priority.
The depth model shown in Fig. 5 (e.g., Monodepth2) illustrates a different case, where changing on-chip capacity alters energy use significantly, but memory-side latency changes little. This suggests that, in such workloads, extra on-chip resources mostly reshape the energy profile rather than reduce the limiting memory-side latency.

\section{Workload Profiling Characterization}
To understand the compute and memory behaviors of XR workloads while avoiding conflating algorithmic bottlenecks with implementation- or mapping-induced limitations, we adopt a two-level profiling methodology. We use NVIDIA Nsight\cite{NsightCompute} to profile GPU executions at the kernel level, quantifying architectural utilization and efficiency (e.g., arithmetic pipeline activity, memory throughput and locality, and scheduling efficiency). Complementarily, we use Intel Pin\cite{luk2005pin} to perform ISA-level instrumentation on the CPU, extracting whole-application characteristics such as instruction composition, memory access patterns, and instruction-level parallelism to capture whole-application instruction-stream signatures (including runtime/control overhead). This dual-perspective workflow helps disentangle workload-facing properties from effects introduced by kernel decomposition, compiler transformations, or mapping choices.
\vspace{-0.8em}
\subsection{GPU Profiling}

We profile all applications using Nsight Compute (NCU) to attribute hardware events to individual kernel launches and aggregate metrics in an execution-time-weighted manner. Experiments are conducted on an NVIDIA Quadro T2000 under the same input resolution and batch=1 setting as Section~3, reflecting real-time XR inference constraints. We emphasize normalized utilization/locality metrics (pct-of-peak, hit rates) to enable cross-kernel and cross-library comparisons. We further aggregate kernels by operator family (e.g., Conv/GEMM, Transform/Resample, Elementwise, Reduce, Memcpy/Data-movement, Search/Index), and analyze a concise set of KPIs capturing compute saturation, bandwidth/locality, and scheduling efficiency (used later in the radar plots of Fig.~6).
\vspace{-0.5em}
\begin{table}[htbp]
\centering
\caption{Main Nsight Compute (NCU) profiling metrics and interpretation used in this work.}
\begin{tabular}{l p{0.8\linewidth}}
\toprule
\textbf{Metric} & \textbf{Interpretation (in our profiling context)} \\
\midrule
\textbf{FMA PIPE\%} &
FMA pipeline utilization (SMSP/SM, normalized to peak). High values indicate strong FMA activity and compute intensity; low values suggest non-FMA--dominated or stall-limited execution. \\

\textbf{SM Util\%} &
SM activity level (fraction of cycles with active execution). With Issue\%, high--low indicates active but stalled execution; high--high indicates sustained issuance. \\

\textbf{L1 Hit\%} &
L1/TEX cache hit rate for global memory accesses. Low values indicate limited near-core reuse or streaming/stride access patterns. \\

\textbf{L2 Hit\%} &
L2 cache hit rate for global memory accesses. High L2 with low L1 suggests reuse captured at LLC; both low indicate poor cacheability or working sets exceeding on-chip capacity. \\

\textbf{DRAM BW\%} &
Sustained external DRAM bandwidth (normalized to peak). High values indicate heavy off-chip traffic and potential bandwidth- or DRAM-service limits, especially with low cache hit rates. \\

\textbf{Issue\%} &
Warp issue efficiency. Low values indicate limited sustained issuance due to stalls (e.g., memory dependencies, divergence, synchronization, or insufficient latency hiding). \\
\bottomrule
\end{tabular}
\end{table}

Among the many hardware counters provided by NCU, we select six concise and representative metrics as key KPIs for architectural analysis shown in Table 2. These KPIs summarize three aspects: compute activity, memory traffic/locality, and issuance efficiency, which provide a compact, widely adopted view for cross-application comparison. We use them as first-order indicators rather than exhaustive stall diagnosis.

\begin{figure}[htbp]
  \centering
  \includegraphics[width=\linewidth]{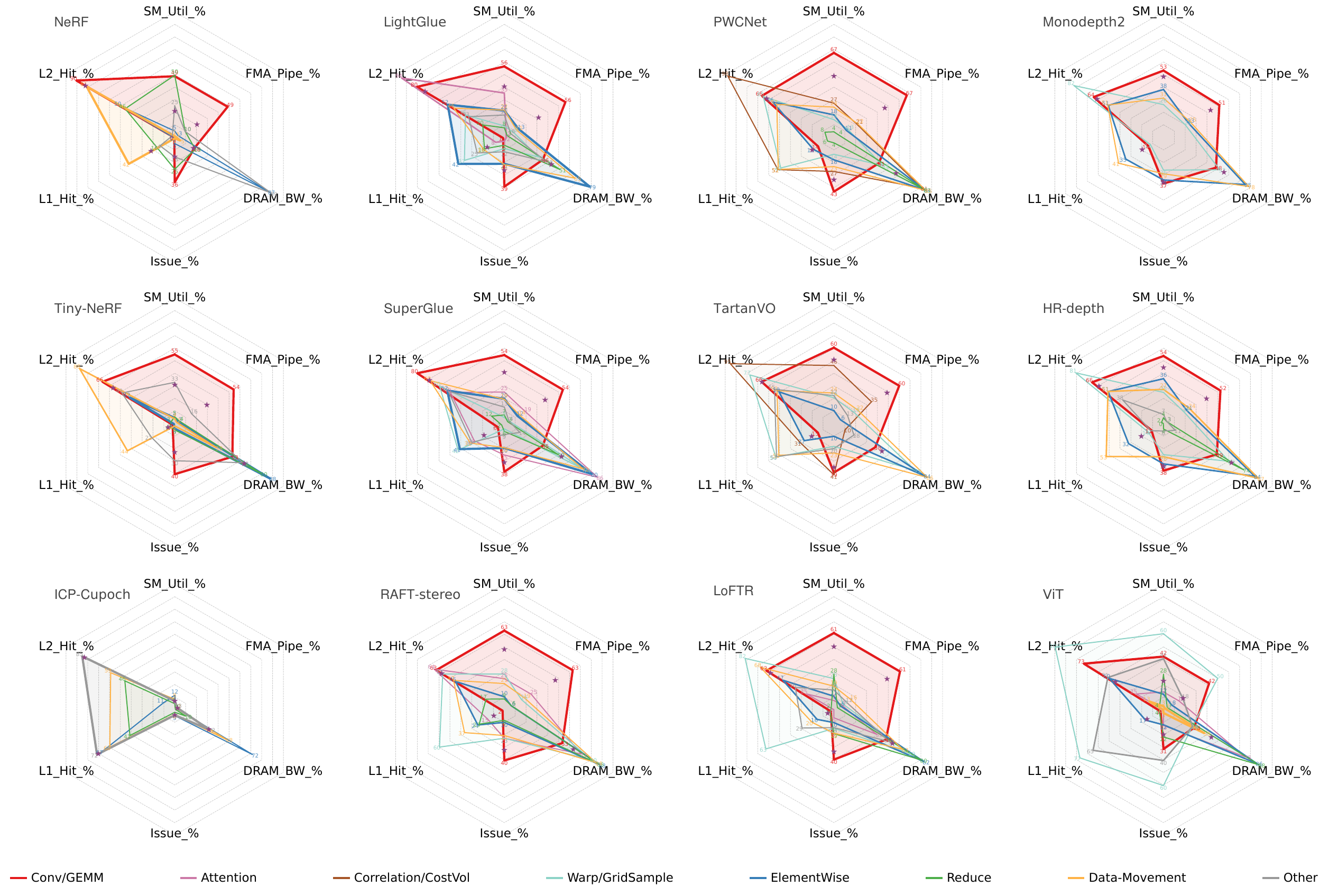}
  \caption{NCU profiling radar plot. Each polygon corresponds to an operator family, with line width proportional to its share of total runtime (thicker lines indicate longer-running families). The star marks the application-level, time-weighted average of each KPI.}

\end{figure}

Fig. 6 presents radar plots of all twelve applications across these six KPIs. Overall, all workloads display moderate compute saturation: the average SM\_Util\% and FMA\_PIPE\% of all the workloads falls in the range of 30-50\% level, which indicates that sustained peak arithmetic throughput is not the dominant limiter for most workloads. By contrast, several operator families exhibit pronounced peaks in DRAM\_BW\%, often accompanied by reduced L1\_Hit\%, elevated L2\_Hit\%, and lower Issue\%. These patterns highlight execution phases where memory traffic and limited instruction throughput become more prominent, in contrast to the moderate compute saturation seen across the workloads overall.

After grouping kernels by operator family across all networks, several recurring trends emerge.
Conv/GEMM (including MLP/FC) generally forms the main compute-leaning backbone: these kernels often show higher SM\_Util\% and FMA\_Pipe\% than other families and frequently account for a large fraction of runtime (thicker curves). Their locality is typically L2-friendly (high L2\_Hit\%), while L1\_Hit\% can be limited—especially in MLP-heavy cases such as NeRF and ViT.
In contrast, bandwidth pressure is most visible in Transform/Resample and related families (e.g., Warp/GridSample, Correlation/CostVol), which often drive high DRAM\_BW\% together with weaker L1 reuse; L2\_Hit\% may still remain high, indicating that L2 captures part of the locality even when substantial misses continue to reach DRAM.
Finally, Elementwise/Reduce/Data-movement kernels act as background structure with lower compute-pipeline activity and varying degrees of bandwidth demand. Reduce kernels often coincide with lower Issue\% (consistent with dependency/synchronization effects), while data-movement kernels impose high traffic with limited reuse (and can be particularly pronounced when arising from index-based moves). In small-footprint workloads such as ICP-Cupoch, high cache hit rates can coexist with low overall utilization, consistent with fragmented execution dominated by short kernels.

Building on the operator-family view, we summarize each application by combining time-weighted average KPIs with the most extreme operator-family behaviors, and identify four recurring execution patterns in how XR workloads interact with the GPU compute and memory hierarchy.

\textbf{Pattern A --- Decoupled, phase-alternating utilization.}
These workloads show moderate application-level SM\_Util\%, FMA\_PIPE\%, and DRAM\_BW\% on average, yet the KPI profiles are clearly pulled in different directions by specific operator families over time.
Conv/MLP phases reach higher FMA\_PIPE\%, while sampling, patchify, and data-movement phases drive up DRAM\_BW\%, often alongside lower Issue\%. L2\_Hit\% often remains high, reflecting that some locality is captured at L2 even as DRAM traffic spikes.
The workload therefore alternates between compute-leaning phases (higher FMA\_PIPE\%) and memory/bandwidth-pressured phases (higher DRAM\_BW\% and lower Issue\%). 
NeRF, TinyNeRF, and ViT fall into this category.

\textbf{Pattern B --- Balanced, cache-friendly pipelines.}
Pattern~B presents a more balanced KPI profile: application-level SM\_Util\%, FMA\_PIPE\%, DRAM\_BW\%, and L1/L2\_Hit\% cluster in a mid-range band, and the dominant Conv/GEMM phases closely match the application averages, indicating that most runtime is spent in regular, compute-leaning stages with stable cache behavior.
While Transform and Reduce families may occasionally create local extremes, their short duration prevents them from shifting the overall profile. As a result, these workloads respond to modest scaling with relatively consistent (though not always linear) improvements.
Monodepth2 and HRDepth are representative of this pattern.

\textbf{Pattern C --- L2-centric, dual-pressure pipelines.}
In Pattern~C, both compute activity and DRAM traffic remain persistently non-trivial.
Application-level SM\_Util\% and FMA\_PIPE\% commonly lie in the 40--50\% range, while DRAM\_BW\% stays elevated but not spiky; meanwhile, L2\_Hit\% is among the highest in the suite, indicating L2 captures a substantial fraction of accesses (and any available temporal locality).
The key differentiator within this pattern is L1\_Hit\%:
(i)~C1 (\emph{severely L1-starved}), where L1\_Hit\% collapses to single digits and primary Conv/GEMM kernels near-zero L1 hits (e.g., RAFT and LoFTR); and
(ii)~C2 (\emph{moderately L1-effective}), where L1\_Hit\% remains non-zero but limited, while execution is still fundamentally L2-dominated (e.g., LightGlue, SuperGlue, PWC-Net, and TartanVO).

\textbf{Pattern D --- Cache-resident, low-intensity workloads.}
Pattern~D sits at the low-utilization / high-hit corner of the KPI space.
Application-level SM\_Util\%, FMA\_PIPE\%, and Issue\% are markedly lower than in the other patterns, while L1/L2\_Hit\% is exceptionally high and DRAM\_BW\% remains moderate, consistent with a small working set and low arithmetic intensity.
Dominant kernels tend to be short and control-heavy (often grouped into \emph{Other}), suggesting that compute and bandwidth resources are over-provisioned relative to the problem size.
Further scaling ALU throughput or DRAM bandwidth is therefore less likely to help, whereas reducing control overhead and improving fast local storage tend to matter more.
Cupoch ICP instantiates this pattern, but the definition is metric-driven and may also capture other small-scale, cache-resident geometric or graph-like workloads.

Notably, we rarely observe a stable “low L2/high L1” mode, nor workloads with persistently low L2 hit rates. This can be attributed to two factors:
First, modern NVIDIA GPUs route nearly all global memory traffic through a unified L2 cache before reaching DRAM; thus, workloads with any moderate temporal or spatial locality tend to achieve reasonable L2 hit rates, and consistently low L2 hits require either very large working sets or highly irregular accesses.
Second, mainstream XR perception kernels—cuDNN-style convolutions, dense GEMM/MLP, standard elementwise—are heavily tiled and prefer coalesced L2 streaming for reuse. By contrast, the per-SM L1 is capacity-limited and shared with shared memory, making high L1 hit rates harder to sustain.
In contrast, large-scale sparse GNNs or recommender models with very irregular access patterns can more easily break cache effectiveness; however, such cloud back-end workloads have different batching and latency constraints and are beyond the XR inference scope of this work.

\subsection{CPU Profiling}

To complement GPU kernel profiling, we profile each application end-to-end under a CPU-only execution setting (i.e., the full pipeline runs on the CPU without any GPU involvement). This view captures whole-application instruction-stream signatures, including control-plane overhead and runtime effects that are invisible in device-side kernel measurements.

Figure~7(a) reports the dynamic instruction mix, showing that arithmetic instructions account for a substantial portion across workloads, while explicit memory operations and control-flow instructions remain non-negligible. To further summarize CPU-side data-access pressure, Figure~7(b) projects each workload onto two complementary ISA indicators: an instruction-level arithmetic intensity
$AI=\frac{\text{arith}\%}{\text{load}\%+\text{store}\%}$ and bytes-per-instruction $BPI$ (bytes accessed per executed instruction). Points are additionally colored by a CPU-side cache hit-rate proxy to reflect cacheability.
Overall, these signatures help distinguish compute-heavy workloads (high $AI$, moderate $BPI$) from memory-pressured workloads (low $AI$ and/or high $BPI$), and highlight cases where poor cacheability correlates with elevated data-movement overhead.
\begin{figure}[htbp]
  \centering
  \newcommand{\FigH}{4.6cm} 

  \begin{minipage}[t]{0.55\linewidth}
    \vspace{0pt}\centering
    \includegraphics[height=\FigH,keepaspectratio]{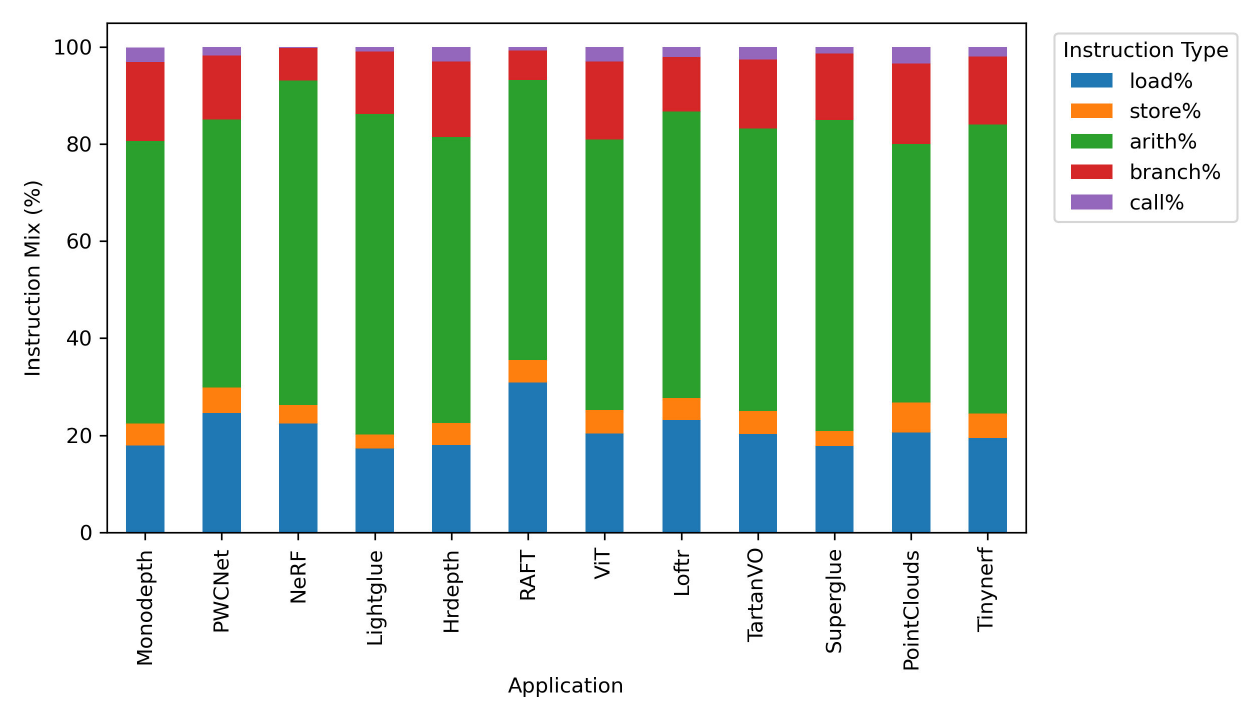}
  \end{minipage}\hfill
  \begin{minipage}[t]{0.45\linewidth}
    \vspace{0pt}\centering
    \includegraphics[height=\FigH,keepaspectratio]{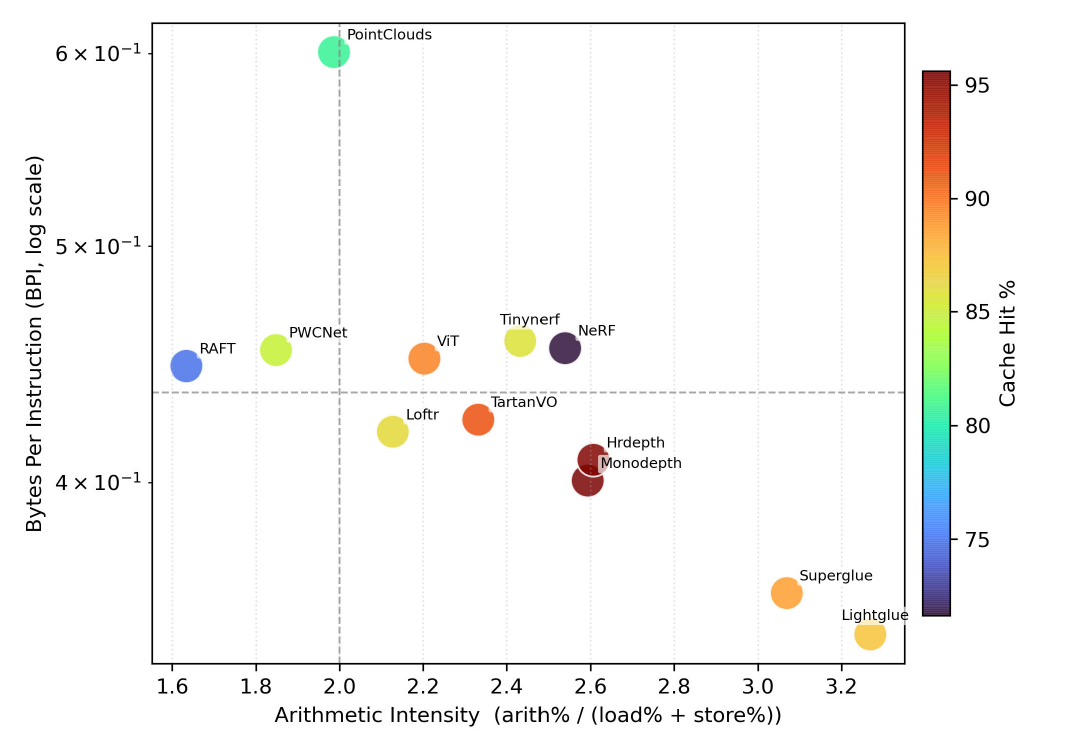}
  \end{minipage}
  \caption{CPU ISA-level profiling results.
(a) Dynamic instruction breakdown (load/store, arithmetic, and control-flow).
(b) AI--BPI projection of each workload, where AI reflects arithmetic vs. explicit memory instruction share and BPI captures bytes accessed per instruction; points are colored by cache hit rate.}
\end{figure}

Based on the AI–BPI position in Fig.~7(b) and the cache-hit rate (CPU-side cacheability proxy), we summarize four recurring regimes in our CPU-side view:

\textbf{Arithmetic-heavy and cacheable (high AI, lower BPI, high hit).}
Workloads in this region exhibit compute-leaning instruction streams with strong cacheability, and are therefore less likely to be dominated by load/store instruction activity. In our suite, HRDepth and MonoDepth fall into this category, with LightGlue and SuperGlue showing a similar signature when included.

\textbf{Mixed/balanced instruction streams (mid AI, mid BPI, mid-to-high hit).}
This regime reflects a mixed structure where both arithmetic and memory instructions are material, yet cacheability remains reasonable. ViT, TartanVO, LoFTR, and TinyNeRF cluster in this region, suggesting a mixed instruction-stream signature rather than a strongly skewed compute- or memory-instruction profile.

\textbf{Memory-instruction–dense (low AI, non-trivial BPI).}
Points in this region indicate higher load/store density relative to arithmetic, typically accompanied by non-trivial data-access demand. PWC-Net and especially RAFT exhibit this signature; RAFT further shows a lower cache-hit rate, compounding memory-instruction pressure with weaker cacheability. This trend is consistent with transform/resampling-heavy stages in these pipelines.

\textbf{Cacheability/data-movement challenged despite decent AI.}
This regime highlights that a compute-leaning instruction mix alone does not guarantee cache-benign behavior: limited reuse or irregular access can still drive cacheability and data-movement pressure even when AI is relatively high. NeRF is the most representative example, showing relatively high AI but noticeably poorer cache-hit rate. PointClouds shows a complementary signature, with substantially higher BPI, indicating much larger per-instruction data-access demand.

\section{Workload Archetypes and Cross-Layer Architectural Implications}
Sections~5 and~6 jointly characterize the architectural behaviors of our XR workload suite from three complementary angles.
In this section, we consolidate these observations into a compact cross-layer behavioral space, then summarize the recurring behaviors into a small set of workload archetypes. Finally, we translate the archetypes into architectural guideline for XR-oriented accelerator design.
\subsection{Cross-Layer Behavioral Space}
To organize the results without overfitting to any single platform or implementation detail, we place each workload in a compact cross-layer space with three complementary dimensions, each tied to a distinct design lever.
\textbf{(D1) On-chip capacity sensitivity (DSE view).}
Our DSE model sweeps L1/LLC capacities under a consistent mapping strategy, quantifying how additional on-chip storage converts potential reuse into realized traffic reduction across the memory hierarchy. This view captures whether a workload exhibits a clear working-set threshold that sharply reduces off-chip pressure, or instead improves gradually with diminishing returns.
\textbf{(D2) GPU execution regime and reuse realization (GPU view).}
GPU profiling characterizes the dominant runtime pressures and where reuse is realized in practice---near the compute units (e.g., at L1/TEX) versus primarily at a larger on-chip level (e.g., L2). In practice, the observed regime is shaped by the dominant operator families, their kernel decomposition, and whether performance/energy pressure manifests as compute throughput limits, on-chip data movement, or off-chip bandwidth demand.
\textbf{(D3) Whole-application instruction pressure and control/memory behavior (CPU view).}
CPU ISA profiling provides a whole-program reference that complements device-kernel-centric GPU profiling. Because the CPU view naturally includes control-plane and runtime effects (e.g., dispatch, synchronization, and framework overhead), it exposes end-to-end instruction pressure and cache-friendliness that are largely invisible when analyzing GPU kernels alone. This helps distinguish workload-facing access/control characteristics from effects that are specific to GPU kernel decomposition and mapping.
Taken together, (D1) captures \emph{capacity leverage} under a unified mapping model, (D2) summarizes \emph{GPU-side execution regimes} and reuse realization in practice, and (D3) provides an \emph{end-to-end reference} for instruction/control pressure and cache behavior. We use these three dimensions as a compact cross-layer space for summarizing XR workload. behaviors.
\subsection{Cross-Layer Workload Archetypes}
We organize archetypes around GPU-side execution regimes, and refine boundaries using DSE capacity trends and CPU ISA overlays. This yields four structural archetypes; phase alternation is treated as a temporal overlay that may co-occur with any structural archetype.

\textbf{Archetype I: Capacity-gated transform pipelines with weak near-cache locality.}
(\emph{Representative applications:} RAFT, PWC-Net, ViT, TartanVO.)
Workloads in this group exhibit a Pattern-C-like GPU profile (high L2 activity and elevated DRAM demand), yet a substantial fraction of their off-chip traffic appears convertible under on-chip scaling within our explored L1/LLC grid. Our DSE sweep shows a pronounced transition band for some workloads (e.g., RAFT): as LLC crosses a working-set boundary, DRAM energy drops sharply and pressure shifts from off-chip traffic toward on-chip reuse. Others (e.g., TartanVO, ViT, and partially PWC) still benefit noticeably, but the transition is more gradual across the LLC axis. This suggests that baseline DRAM demand is largely driven by capacity misses and incomplete reuse capture rather than purely compulsory streaming. ViT primarily falls into this regime due to its capacity-gated attention stages, while its end-to-end pipeline can also exhibit phase alternation.

\textbf{Archetype II: Flat-response transform/matching pipelines with hard-to-realize reuse.}
(\emph{Representative applications:} LightGlue, SuperGlue, LoFTR.)
These workloads show a Pattern-C-like GPU profile, yet their DSE curves are markedly flat: across the explored L1/LLC budgets, DRAM remains the dominant energy component. CPU-side ISA signatures are relatively compute-leaning with reasonably high cache hit rates, indicating that algorithm-level memory intensity alone does not fully explain the persistent DRAM demand on GPU. Instead, the evidence is consistent with locality potential that is difficult to materialize on the GPU, due to factors such as indirect indexing (gather/scatter), intermediate materialization across stages, and overhead from fragmented kernel execution. As a result, cache scaling alone yields limited returns; effective support requires explicit staging and lower-overhead data-movement primitives.

\textbf{Archetype III: Balanced, cache-friendly pipelines with diminishing returns under cache scaling.}
(\emph{Representative applications:} Monodepth2, HRDepth; and the MLP compute stage in NeRF-style networks.)
These pipelines map well to conventional dense acceleration (Conv/GEMM/MLP-dominant), and once modest on-chip capacity captures the dominant working sets, further cache scaling yields diminishing returns. This regime is largely covered by conventional dense-acceleration designs: further generic scaling can still improve absolute performance, but its marginal value is reduced in XR SoC settings where area/power budgets are constrained and other archetypes expose more acute bottlenecks. For NeRF-style workloads, we emphasize that this archetype describes the \emph{MLP compute stage} in isolation; the end-to-end pipeline can still exhibit phase alternation due to the rendering/sampling loop (temporal overlay).

\textbf{Archetype IV: Irregular geometric kernels that are overhead- and latency-sensitive.}
(\emph{Representative application:} ICP-Cupoch type point-cloud workload.)
These workloads often fail to sustain high GPU utilization not because of peak arithmetic throughput, but because execution is dominated by irregular traversal and short dependency chains---e.g., repeated nearest-neighbor queries, gather/scatter-style indexing, and fine-grained synchronization across iterative refinement. On the CPU, these kernels often manifest as data-movement--heavy instruction streams with lower arithmetic density; cache hit rates can vary with the data structure and working-set size, but the behavior is consistent with irregular spatial traversal and fine-grained synchronization (e.g., pointer-chasing-style access in spatial data structures). Consequently, simply provisioning larger caches or higher peak bandwidth does not directly resolve the dominant limitation; instead, effective support requires low-overhead irregular access primitives, latency-tolerant scheduling, and explicit staging mechanisms that reduce synchronization and intermediate materialization.

\textbf{Temporal overlay: Phase-alternating rendering/sampling pipelines.}
(\emph{Representative applications:} NeRF-style networks, TinyNeRF, ViT.)
Beyond the structural archetypes, several pipelines exhibit phase alternation: execution repeatedly switches between compute-leaning phases (e.g., MLP inference/shading) and access-/sampling-heavy phases (e.g., ray marching, resampling/warping, patchify, gather-heavy indexing). Even when additional cache capacity reduces off-chip traffic for a given pipeline, end-to-end efficiency can still be governed by a mismatch in the optimal compute--bandwidth--capacity balance across phases. Consequently, a single static provisioning point is inherently suboptimal for at least one phase, motivating phase-aware scheduling and elastic resource allocation.

\subsection{Architectural Implications}
Sections~5--6 show that XR workloads are rarely governed by a single resource; instead, outcomes are shaped by the interaction between working-set fit, reuse realization, and execution structure over time. On-chip capacity is therefore not a universal ``more cache'' knob: it is highly effective when it crosses a working-set threshold, but yields diminishing returns when reuse is hard to materialize near compute or when overhead and phase shifts dominate. Across our suite, three mechanisms recur: capacity thresholds, reuse-realization gaps, and temporal/overhead sensitivity. We summarize actionable architectural implications along these mechanisms below.

\textbf{Implication 1: Provision enough last-level on-chip capacity to cross the working-set threshold when it exists.}
For capacity-gated regimes, sufficient last-level on-chip capacity is the first-order requirement to avoid repeated spill/fill and persistent DRAM penalties. Below the threshold, scaling compute alone is unlikely to change the fundamental off-chip cost profile.

\textbf{Implication 2: When near-cache reuse is structurally hard, expose explicit locality/staging mechanisms.}
For flat-response workloads (Archetype~II), passive caching alone cannot reliably convert off-chip traffic into on-chip reuse. Architectures should expose scratchpad-style staging (e.g., programmable buffering and double-buffering) and lightweight gather/scatter-friendly data-movement support so that runtimes can reduce intermediate materialization, coalesce irregular accesses, and amortize synchronization across phases.

\textbf{Implication 3: Provision elasticity and reallocate budgets across phases and diminishing-return regimes.}
For phase-alternating pipelines, static provisioning inevitably underutilizes at least one phase; for balanced cache-friendly regimes, returns also diminish once dominant working sets are on-chip. A practical XR SoC strategy is therefore to avoid over-investing in generic scaling for well-served phases, and instead allocate area/power toward elastic execution support (phase-aware scheduling and adaptable on-chip dataflow) and toward regimes where additional resources qualitatively change the cost profile (capacity thresholds and irregularity tolerance).

\textbf{Implication 4: For irregular and overhead-sensitive kernels, prioritize low-overhead execution support over raw bandwidth.}
In latency/irregularity-dominated regimes, the critical need is to reduce overhead and tolerate irregularity, rather than only increasing bandwidth. Lightweight execution modes, irregular-aware scheduling, efficient indexing primitives, and low-latency synchronization close to on-chip data can help mitigate kernel fragmentation and dependency-chain stalls in throughput-oriented execution.

\section{Conclusion}

In this paper, we present a systematic architecture-level characterization of representative edge XR workloads through three complementary views. We combine: (1) a lightweight, fast analytical DSE model to systematically sweep on-chip memory capacities and mapping choices; (2) cross-platform profiling—GPU kernel-level and CPU ISA-level—to quantify utilization, locality, scheduling efficiency, and end-to-end instruction-stream characteristics. Together, these views form an “architectural cartography” of XR behaviors, enabling comparable analysis of working-set pressure, bandwidth demand, compute utilization, and runtime overhead across 12 representative XR kernels. Based on the resulting cross-layer behavioral space, we distill five reproducible workload archetypes and translate them into actionable architectural implications for next-generation XR SoCs.

Our results indicate that XR performance and energy efficiency are often governed by threshold and phase effects rather than by linear gains from generic resource scaling. Several kernels exhibit sharp on-chip capacity tipping points, a reuse-realization gap where algorithmic reuse exists but is poorly captured by passive caching, and temporal variability induced by multi-stage graphs. Consequently, XR SoC design should emphasize threshold- and phase-driven organization: provision on-chip capacity where it unlocks capacity-gated phases; add explicit staging support and irregular datapaths for cache-resistant phases; adopt phase-aware scheduling and elastic resource allocation for phase-alternating pipelines; and, for overhead-dominated kernels, prioritize reducing synchronization and scheduling overheads over further scaling peak compute or bandwidth.

However, there are some remaining assumptions. First, we intentionally focus on memory-hierarchy sensitivity and cross-workload comparability rather than cycle-accurate or implementation-calibrated prediction. Compute energy and latency are modeled using simplified proxies (e.g., FLOP-proportional compute energy and a roofline-style latency treatment), and the SIMD architectural template is used strictly as a unified analysis baseline rather than as an optimal XR accelerator proposal. Accordingly, our DSE outcomes should be interpreted as trend- and tipping-point–oriented sensitivity evidence, not as absolute projections for a specific silicon implementation. Second, our analysis targets single-application behavior under batch=1 near-real-time inference constraints and does not model full-system runtime effects such as multi-application concurrency, interference, or dynamic interconnect/runtime policies.

\bibliographystyle{ACM-Reference-Format}
\bibliography{reference}
\end{document}